\documentclass[
    ,final            
  ]
  {aipproc}

\layoutstyle{6x9}

\begin{document}

\title{The nrMSSU(5) and universality of soft masses}


\author{F. Borzumati}{
  address={Department of Physics, National Taiwan University,
          Taipei 10617, Taiwan}
}

\author{T. Yamashita}{
  address={Department of Physics, Nagoya University,
          Nagoya 464-8602, Japan}
}

\begin{abstract}
We discuss the problem of universality of the soft,
 supersymmetry-breaking terms in the minimal supersymmetric SU(5) 
 model~(MSSU(5)) completed with flavor-dependent nonrenormalizable 
 operators~(NROs), or nrMSSU(5). 
These are exploited to correct the wrong fermion spectrum and to slow
 down the too-fast decay rate of the proton that the MSSU(5) model
 predicts. 
In general, the presence of such operators in the superpotential and 
 K\"ahler potential gives rise to tree-level flavor- and CP-violating
 entries in the sfermion mass matrices at the cutoff scale, even when
 the mediation of supersymmetry (SUSY) breaking is generation and
 field-type independent. 
We identify the conditions under which such terms can be avoided.  
\end{abstract}

\maketitle


The MSSU(5) model is plagued by two fundamental problems. 
It predicts a wrong fermion spectrum and a too-large rate for the decay
 of the proton.
The first is clearly encoded in the matter superpotential:  
\begin{equation}  
 W^{\rm MSSU(5)}_{\rm M} =   
   \sqrt{2} \, \bar{5}_M  Y^5    \, 10_M \bar{5}_H  
 - \frac{1}{4} 10_M       Y^{10} \, 10_M 5_H  .
\label{eq:WminSU5matt}
\end{equation}
Different matter multiplets of the Standard Model (SM), such as
 $D^c$ and $L$, collected in the same MSSU(5) multiplet $\bar{5}_M$,
 share the same Yukawa coupling $Y^5$, whereas two distinct couplings
 are present in the SM.
Surprisingly, the mass predictions of the MSSU(5) model go wrong only
 for the first and second generation, but not for the third. 
Thus, the corrections needed to fix the fermion spectrum are small and
 can be provided by NROs. 
These in general exist in any model and their effect can be quite
 tangible if the cutoff of the model, in this case the Planck mass
 $M_{\rm P}$, is not too far away from the scale of the model itself, in
 this case the grand-unification scale. 
A general class of NROs that split the masses of first and second
 generations down quarks and charged leptons is given by:
\begin{equation}
 Op^{5}         = 
 \sum_{n+m=0}^k\,
 \sqrt{2}\ \bar{5}_M \,C_{n,m}^{5} 
 \left(\!\displaystyle{\frac{24_H^T}{M_{\rm P}}}\!
 \right)^{\!\!n}  \! 10_M 
 \left(\!\displaystyle{\frac{24_H}{M_{\rm P}}}\!
 \right)^{\!\!m}  \! \bar5_H,
\label{eq:NRO5}  
\end{equation}
 where the SU(5) generators act on the antifundamental rather than 
 the fundamental representation.
This class contains already the first term of
 Eq.~(\ref{eq:WminSU5matt}).
Indeed, the coefficient  $C_{o,o}^{5}$ in the first term in $Op^{5}$ is
 equal to $Y^5$.
Notice that the suppression factors $1/(M_{\rm P})^n$ in this class can
 be partially compensated by the scalar {\it vev} $v_{24}$ of the
 adjoint field $24_H$, giving rise to the much milder suppression
 factors $s^n$, with $s= v_{24}/M_{\rm P}$, of ${\cal O}(10^{-2})$.
Once $24_H$ acquires this SU(5)-breaking {\it vev}, the four SM
 operators in which the first term in Eq.~(\ref{eq:WminSU5matt})
 decomposes, are what $Op^{5}$ reduces to, up to negligible terms: 
\begin{equation}
  -D^c {\bf Y}^{5}_{D\,} Q H_d,      \quad\,
  -E^c ({\bf Y}^{5}_{E\,})^T L H_d,   \quad\,
  -D^c {\bf Y}^{5}_{DU} U^c H_D^C,   \quad\,
  -L\, {\bf Y}^{5}_{LQ} Q H_D^C . 
\label{eq:5effSMdecomp}
\end{equation}
Their coefficients ${\bf Y}^{5}_i$, with 
 $(i={\scriptstyle D,E,DU,LQ})$, however, are not equal to $Y^5$
 anymore, but are effective couplings given by:
\begin{equation}
  {\bf Y}^{5}_i                           \ = \ 
  \sum_{n+m=0}^k \!
  {\bf Y}_i^{5}\vert_{n+m}                   \ = \
  \sum_{n+m=0}^k 
   C_{n,m}^{5} \, s^{(n+m)}  
   \left(\left(I_{\bar{5}_M}\right)_i\right)^n
   \left(\left(I_{\bar{5}_H}\right)_i\right)^m,  
\label{eq:effY5def}
\end{equation}
 and contain enough freedom in the coefficients $C_{n,m}^{5}$ to give
 rise to the correct fermion spectrum.
(The quantities $(I_{\bar{5}_M})_i$ and $(I_{\bar{5}_H})_i$ are hypercharge
 factors~\cite{BY}.)

We shall not touch upon the second problem of the MSSU(5) model, that of
 proton decay. 
We remind only that a solution to the problem can also be provided by
 NROs~(~\cite{BY},~\cite{BMYtalks}, and references therein).

Thus, flavor-violating NROs induce corrections to tiny quantities
 concerning mainly the first two generations. 
Because of this, the repercussion that such NROs have on the usual
 evaluation of flavor violations in the sfermion sector (sFVs) was for
 a long time (and still is) considered negligible.
These sFVs are induced at the quantum level by Yukawa couplings of 
 third-generation quarks and possibly seesaw neutrino, when starting
 from a structure, breaking SUSY softly, which is free at the cutoff
 scale from such violations at the tree level.
A large body of work, indeed, exists that analyzes the simple
 correlation between flavor violations induced by right-handed seesaw
 neutrinos in the slepton sector~(sLVs)\cite{NuY->sLFV} and in the
 squark sector (sQVs), in particular in the sector of the right-handed
 down squarks.
This correlation was first noticed in Ref.~\cite{NuY->sQFV}.

Several papers have appeared by now on the effect of NROs on flavor
 violations. 
They differ in the way NROs are treated. 
Nevertheless, they agree on the fact that once NROs are included 
 the simplicity of the above mentioned correlations between sLFVs and
 sQFVs due to seesaw-neutrinos interactions can be considerably spoiled,
 even when starting at the cutoff scale with soft terms that are free
 from such violation at the tree level.
This holds also for second-third generation flavor
 transitions~\cite{ParkEt,BY}.

One crucial assumptions of all studies of radiatively induced sFVs,
 is that of universality of soft terms at the cutoff scale.
With this expression it is commonly understood that
 the matrices of soft mass squared are  generation and field-type
 independent, {\it i.e.} are all proportional to the unit matrix with
 the same proportionality constant; 
 that the matrices of trilinear and bilinear soft parameters are
 aligned with, or  proportional to, the corresponding superpotential
 Yukawa couplings and  mass terms.
The proportionality constant for the two sets of bilinear and trilinear 
 soft terms are, in general, assumed to be different one from the other,
 and also different from those of the soft mass squared.
This concept of universality implies a flavor-blind mechanism of 
 mediation of SUSY breaking, {\it i.e.} that the couplings of the
 field {\sf X} mediating this breaking to the various terms in the
 superpotential and K\"ahler potential is independent of generations and
 field-types.
It turns out that for nonvanishing NROs, the above criterium is not
 sufficient to guarantee vanishing tree-level sFVs at the cutoff
 scale.

We give in the following a simple example to illustrate this point. 
Since the field $24_H$ has a nonvanishing auxiliary {\it vev} $F_{24}$, 
 the soft trilinear terms corresponding to the Yukawa operator in 
 Eq.~(\ref{eq:NRO5}) are obtained not only from:
\begin{equation}
  Op^{5}({\sf{X}})  = 
 \sum_{n+m=0}^k
   \left(\!\frac{{\sf{X}}}{M_{\rm P}}\!\right)        
   \sqrt{2} \,  \bar{5}_M \,  a_{n+m}^{5} C_{n,m}^{5} 
   \left(\!\frac{24_H^T}{M_{\rm P}}\!\right)^{\!n}  \! {10}_M 
   \left(\!\frac{24_H}{M_{\rm P}}\!\right)^{\!m}  \! \bar5_H ,
\label{eq:NRO5tildeX}
\end{equation}
 once {\sf X} is replaced by its auxiliary component, $F_{\sf{X}}$, but also 
 by the operators $Op^{5}$ themselves, when one $24_H$ is replaced by 
 its auxiliary {\it vev} $F_{24}$.
Notice that the conventional notion of universality was already assumed in 
 the previous equation, that is, the various operators of same 
 dimensionality, as those with coefficients 
 $C_{1,0}^{5}$ and $C_{0,1}^{5}$, or with coefficients 
 $C_{2,0}^{5}$, $C_{0,2}^{5}$ and $C_{1,1}^{5}$ were assigned the same constants 
 $ a_{1}^{5}$ and  $ a_{2}^{5}$, respectively.

The effective trilinear operators obtained in this case: 
\begin{equation}
-\widetilde{D}^c { \bf A}^{5}_{D\phantom u}
 \widetilde{Q} H_d,                                        \quad\,
-\widetilde{E}^c({\bf A}^{5}_{E\phantom q})^T
 \widetilde{L} H_d,                                        \quad\,
-\widetilde{D}^c {\bf A}^{5}_{DU}
 \widetilde{U}^c H_D^C,                                    \quad\,
-\widetilde{L}\, {\bf A}^{5}_{LQ}
 \widetilde{Q} H_D^C ,   
\label{eq:tilde5effSMdecomp}
\end{equation}
 have coefficients ${\bf A}^{5}_i$, $(i={\scriptstyle D,E,DU,LQ})$
 given by:
\begin{equation}
 {\bf A}^{5}_i  = \!  
 \sum_{n+m=0}^k \!
  {\bf A}_i^{5}\vert_{n+m}                 = \!
  \displaystyle{\sum_{n+m=0}^k} \!
   \left[ A_{n+m}^{5}+\!\left(n\!+\!m\right)f_{24}
   \right] C_{n,m}^{5} \, s^{(n+m)}  
   \left(\left(I_{\bar{5}_M}\right)_i\right)^n \!
   \left(\left(I_{\bar{5}_H}\right)_i\right)^m \!,
\label{eq:effA5def}
\end{equation}
 where the massive parameters $A_{n+m}^{5}$ are
 $A_{n+m}^{5} = a_{n+m}^{5} ({F_{\sf{X}}}/{M_{\rm P}})$, and 
 $f_{24}$ is the ratio $f_{24} = {F_{24_H}}/{v_{24}}$.
The effective trilinear couplings ${\bf A}^{5}_i $ are 
 not aligned with the effective couplings ${\bf Y}^{5}_i$,
 not even if the various coefficients $a_{n+m}^{5}$ are 
 taken to be equal for all values of $n+m$.

The superpotential that gives rise to above effective 
 trilinear terms has the form:
\begin{equation}
\begin{array}{lll}
 W  &\!\! = \!\!&
   M_{ij}  \left(1 + b f_X \theta^2\right)\phi_i\phi_j 
\\[1.1ex]       & & 
\hspace*{-0.5truecm}
  +Y_{ijk} \left(1 + a f_X \theta^2\right)\phi_i\phi_j\phi_k
  +(C_{ijkl}/M_{\rm P})\left(1 +a_1 f_X\theta^2\right)\phi_i\phi_j\phi_k\phi_l
  +\cdots,
\label{eq:Wuniv}
\end{array}
\end{equation}
 where  $M$ is one of the superheavy Higgs masses $M_5$ or $M_{24}$,
 or a seesaw massive parameter, $Y$ any Yukawa coupling, and $C$ any
 coupling of a NRO of dimension five.
We have changed here our notation in order to simplify our equations:
 the fields $\phi_i$ in these superpotentials are SU(5) multiplets, and
 the field indices denote field types and generations. 
For the problem of tree-level sFVs in the effective soft mass squared,
 which also exist, in general, when the conventional notion of
 universality is implemented, we refer the reader to Ref.~\cite{BY}.
It turns out that sFVs in the superpotential can be avoided if
 $\sf{X}$ couples to its terms in one of the two factorizable forms:
\begin{equation}
 W              =
 \left(1 + a f_X \theta^2\right)
 \left[
     M_{ij}  \phi_i\phi_j 
    +Y_{ijk}\, \phi_i\phi_j\phi_k
  +  (C_{ijkl}/M_{\rm P})\phi_i\phi_j\phi_k\phi_l +\cdots \right],
\label{eq:WunivSTR1}
\end{equation}
\begin{equation}
\begin{array}{lll}
 W              &=&
  \left(1 +a^\prime f_{\sf X} \theta^2\right)
\left[
   M_{ij}  \left(1 -2 f_{24} \theta^2\right)\phi_i\phi_j 
  +Y_{ijk} \left(1 -3 f_{24} \theta^2\right)\phi_i\phi_j\phi_k
\right.
\\[1.1ex]       & & 
\left.
\phantom{
  \left(1 +a^\prime f_{\sf X} \theta^2\right)}
 + (C_{ijkl}/M_{\rm P})
 \left(1 -4 f_{24} \theta^2\right)\phi_i\phi_j\phi_k\phi_l
  +\cdots
\right].
\label{eq:WunivSTR3}
\end{array}
\end{equation}
The first form of the superpotential is not stable under field
 redefinitions, which reduce it to the second form.
We have reported it, however, because it can be shown that sFVs in the
 K\"ahler potential can be also simultaneously removed if, in the basis
 of Eq.~(\ref{eq:WunivSTR1}), ${\sf{X}}$ couples to the  K\"ahler
 potential also in a special factorizable way:
\begin{equation}
 K({\sf X}^\ast,{\sf X}) = 
\Big[ 
 1 - \delta_K      f_{\sf X} \theta^2 
   - \delta_K^\ast f_{\sf X}^\ast \bar{\theta}^2 
   + \left(\vert d \vert^2
   + \left\vert \delta_K\vert^2\right) \right\vert f_{\sf X} \vert^2
     \theta^2 \bar{\theta}^2
\Big] K ,
\label{eq:KAELERstrongUNIV}
\end{equation} 
 where $K$ is the SUSY-conserving part of the K\"ahler potential:
\begin{eqnarray}
 K =  
  \phi^\ast_i \phi_i
 -\frac{(d_5)_{ijk}}{M_{\rm P}} \phi^\ast_i \phi_j \phi_k  
 -\frac{(d_6)_{ijkh}}{M_{\rm P}^2}
              \phi^\ast_i \phi_j^\ast \phi_k \phi_h
 -\frac{(d_6^\prime)_{ijkh}}{M_{\rm P}^2}
              \phi^\ast_i \phi_j \phi_k \phi_h + {\rm H.c.} 
+ \cdots,  
\label{eq:KAELERnoX}
\end{eqnarray}
 with coefficients $d_5$, $d_6$, and $d_6^\prime$ all flavor dependent. 
This form shows how the various field redefinitions reducing the
 SUSY-conserving part $K$ to its canonical form
 eliminate also the corresponding flavor dependent NROs in the
 SUSY-breaking part.

It is now easy to show that, by neglecting terms of ${\cal O}(s^3)$ and
 tiny terms of type $\phi^* \phi^* \phi \phi$, and by making use of
 SUSY-breaking field redefinitions, it is possible to bring 
 $ K({\sf X}^\ast,{\sf X})$ to the special form: 
\begin{equation}
 K({\sf X}^\ast,{\sf X}) \simeq 
\Big[ 1 +\vert d \vert^2 \vert f_{\sf X}\vert^2
                         \theta^2 \bar{\theta}^2
\Big]
\left(\delta_{ih} - s^2 (d_6)_{ijjh}\right) \phi^\ast_i \phi_h ,
\label{eq:KpotALMOSTmin} 
\end{equation}
 and the superpotential to the form in
 Eq.~(\ref{eq:WunivSTR3}).
A further SUSY-conserving, but SU(5)-breaking field
 redefinition, yields:
\begin{equation}
 K({\sf X}^\ast,{\sf X}) \simeq 
\Big[ 1 +\vert d \vert^2 \vert f_{\sf X}\vert^2
                         \theta^2 \bar{\theta}^2
\Big]
  \phi^\ast_i \phi_i .
\label{eq:KpotMIN} 
\end{equation}
We obtain then a realistic MSSU(5) model with NROs in the superpotential
 in the form of Eq.~(\ref{eq:WunivSTR3}), and with effective soft masses
 squared equal to  $\widetilde{m}^2$,
 $ \widetilde{\bf m}_{i}(M_{\rm P}) = \widetilde{m}\, {\mathop{\bf 1}}$,
 for any field $i$; trilinear effective couplings aligned with the
 effective Yukawa couplings, as for example in:
 $\mbox{\bf A}^{5}_i(M_{\rm P}) = A_0 \mbox{\bf Y}^{5}_i $,
 $(i={\scriptstyle D,E,DU,LQ})$, with $A_0 = A_0^\prime - 3 f_{24}$;  
 $B$ parameters expressed in terms of the corresponding superpotential
 massive parameter as ${\bf B}_i (M_{\rm P}) =  B_0  \mbox{\bf M}_i $, 
 with: $ B_0 = A_0^\prime - 2 f_{24}$.
That is, we recover the type of universality usually advocated for models
 without NROs, but for effective couplings, or the type of universality 
 emerging in the flat limit from supergravity models with a sequestered
 hidden sector. 
The three free parameters to be specified in addition to the gaugino mass
 can be chosen to be $\widetilde{m}$, $A_0$, $B_0$. 
In this case, $f_{24}$ is fixed in terms of $B_0$ and $A_0$,
 $f_{24} = B_0 - A_0$. 

With this choice of universality, the arbitrariness induced by NROs
 remains confined to the Yukawa sector. As mentioned earlier, this
 does modify the MSS(5) model predictions for sFVs and in particular
 the correlations between sLFVs and sQFVs.

\vspace*{0.5truecm}
\noindent This work is supported in part by the Research Project of NTU,
 Taiwan, grant No 97R0066-60 (F.B.) and JSPS (T.Y.).

\end{document}